\documentclass[prd,aps,preprint,eqsecnum,floats,showpacs]{revtex4}
\begin{document}
\title{ The gluonic condensate from the hyperfine splitting  $M_{\rm cog}(\chi_{cJ})-M(h_c)$ in charmonium}
\author{A.M. Badalian}
\affiliation{State Research Center, Institute of Theoretical and
Experimental Physics, Moscow, Russia}
\author{B.L.G. Bakker}
\affiliation{Vrije Universiteit, Department of Physics,
Amsterdam, The Netherlands}
\date{\today}
\newcommand{\beq}{\begin{eqnarray}}
\newcommand{\eeq}{\end{eqnarray}}
\newcommand{\be}{\begin{equation}}
\newcommand{\ee}{\end{equation}}
\def\la{\mathrel{\mathpalette\fun <}}
\def\ga{\mathrel{\mathpalette\fun >}}
\def\fun#1#2{\lower3.6pt\vbox{\baselineskip0pt\lineskip.9pt
\ialign{$\mathsurround=0pt#1\hfil ##\hfil$\crcr#2\crcr\sim\crcr}}}
\newcommand{\veX}{\mbox{\boldmath${\rm X}$}}
\newcommand{{\SD}}{\rm SD}
\newcommand{\pp}{\prime\prime}
\newcommand{\veY}{\mbox{\boldmath${\rm Y}$}}
\newcommand{\vex}{\mbox{\boldmath${\rm x}$}}
\newcommand{\vey}{\mbox{\boldmath${\rm y}$}}
\newcommand{\ver}{\mbox{\boldmath${\rm r}$}}
\newcommand{\vesig}{\mbox{\boldmath${\rm \sigma}$}}
\newcommand{\vedelta}{\mbox{\boldmath${\rm \delta}$}}
\newcommand{\veP}{\mbox{\boldmath${\rm P}$}}
\newcommand{\vep}{\mbox{\boldmath${\rm p}$}}
\newcommand{\veq}{\mbox{\boldmath${\rm q}$}}
\newcommand{\vez}{\mbox{\boldmath${\rm z}$}}
\newcommand{\veS}{\mbox{\boldmath${\rm S}$}}
\newcommand{\veL}{\mbox{\boldmath${\rm L}$}}
\newcommand{\veR}{\mbox{\boldmath${\rm R}$}}
\newcommand{\ves}{\mbox{\boldmath${\rm s}$}}
\newcommand{\vek}{\mbox{\boldmath${\rm k}$}}
\newcommand{\ven}{\mbox{\boldmath${\rm n}$}}
\newcommand{\veu}{\mbox{\boldmath${\rm u}$}}
\newcommand{\vev}{\mbox{\boldmath${\rm v}$}}
\newcommand{\veh}{\mbox{\boldmath${\rm h}$}}
\newcommand{\verho}{\mbox{\boldmath${\rm \rho}$}}
\newcommand{\vexi}{\mbox{\boldmath${\rm \xi}$}}
\newcommand{\veta}{\mbox{\boldmath${\rm \eta}$}}
\newcommand{\veB}{\mbox{\boldmath${\rm B}$}}
\newcommand{\veH}{\mbox{\boldmath${\rm H}$}}
\newcommand{\veE}{\mbox{\boldmath${\rm E}$}}
\newcommand{\veJ}{\mbox{\boldmath${\rm J}$}}
\newcommand{\veal}{\mbox{\boldmath${\rm \alpha}$}}
\newcommand{\vegam}{\mbox{\boldmath${\rm \gamma}$}}
\newcommand{\vepar}{\mbox{\boldmath${\rm \partial}$}}
\newcommand{\llan}{\langle\langle}
\newcommand{\rran}{\rangle\rangle}
\newcommand{\lan}{\langle}
\newcommand{\ran}{\rangle}
\begin{abstract}
The precision measurement of the hyperfine splitting $\Delta_{\rm HF}
(1P, c\bar c)=M_{\rm cog} (\chi_{cJ}) - M(h_c) = -0.5 \pm 0.4$ MeV in
the Fermilab--E835 experiment allows to determine the gluonic
condensate $G_2$ with  high accuracy if the gluonic correlation length
$T_g$ is fixed. In our calculations the negative value of $\Delta_{\rm
HF} = -0.3 \pm 0.4$ MeV is obtained only if the relatively small $T_g =
0.16$ fm and $G_2 = 0.065 (3)$ GeV${}^4$ are taken.  These values
correspond to the ``physical'' string  tension $(\sigma \approx 0.18 $
GeV$^2$).  For $T_g \ge 0.2$ fm the hyperfine splitting is positive and
grows for increasing $T_g$. In particular for $T_g = 0.2$ fm and $G_2 =
0.041 (2)$ GeV${}^4$ the splitting $\Delta_{\rm HF} = 1.4 (2)$ MeV is
obtained, which is in accord with the recent CLEO result.
\end{abstract}
\maketitle
\section{Introduction}
Recently, the CLEO Collaboration has presented preliminary results on
the mass of the $h_c(1^1P_1)$ resonance in the reaction $\psi'\to
\pi^0h_c$ \cite{ref.1}. It appeared that the central value of the
observed mass, $M(h_c) = 3524.4\pm 0.9$ MeV, is about 1.5 MeV lower
than the one found in the recent Fermilab--E835 experiment, in the
$p\bar p\to h_c \to \eta_c\gamma\to 3\gamma$ reactions, where $M(h_c)
=3525.8 \pm 0.4$ MeV is measured \cite{ref.2}. Consequently, the
central values of the hyperfine (HF) splittings, $\Delta_{\rm HF} (1P,
c\bar c)=M_{\rm cog} (\chi_{cJ})-M(h_c)$, corresponding to the measured
masses $M(h_c)$, have different signs:
\beq
 \Delta^{(1)}_{\rm HF} (\exp) & = & -0.5\pm 0.4~{\rm MeV\;\;(E835)},
\label{eq.1} \\
 \Delta^{(2)}_{\rm HF} (\exp) & = & +0.96\pm 1.03 {\rm MeV\;\; (CLEO)},
\label{eq.2}
\eeq
although they are compatible within $2\sigma$.  (In Eqs.~(\ref{eq.1},
\ref{eq.2}) the value $M_{\rm cog}(\chi_{cJ})= 3525.32\pm 0.13$ MeV is
taken from PDG \cite{ref.3}). The small numerical difference between these
experimental values makes an essential difference for the theoretical
interpretation, as we will show in this paper.

Already many years ago it was understood that the sign and small
magnitude of $\Delta_{\rm HF} (1P)$ in charmonium occur due to the
almost equal (and small) contributions from the perturbative (P) and
nonperturbative (NP) HF interactions \cite{ref.4}:
\be
 \Delta_{\rm HF} (1P) =\Delta^P_{HP} (1P) + \Delta^{\rm NP}_{HF }
 (1P),
\label{eq.3}
\ee
where the perturbative contribution is always negative
\cite{ref.5,ref.6}. Just due to the cancellation between the negative
perturbative contribution and the positive NP one, the value
$\Delta_{\rm HF} (1P, c\bar c) \cong - 1$ MeV has been calculated in
\cite{ref.6} for a value of the gluonic condensate $G_2 \cong 0.042$
GeV$^4$, while in Ref.~\cite{ref.7} the same HF splitting has been
obtained for the significantly smaller value $G_2 \approx 0.02$
GeV$^4$. The reason behind this difference will be explained below and
comes from the fact that the gluonic condensate actually  enters
$\Delta^{\rm NP}_{\rm HF}(1P)$ in the combination $G_2 T^2_g$ ($T_g$ is
the gluonic correlation length) and therefore the extracted value of
$G_2$ depends also on the correlation length $T_g$ used. Unfortunately,
at  present there is a large uncertainty in the value of the gluonic
condensate, even in the framework of the same approach, like QCD sum
rules. Values ranging from $G_2=0.012$ GeV$^4$ up to $G_2\cong 0.07 (3)$
GeV$^4$ are used \cite{ref.8,ref.9}.

The new experimental value of the $h_c$ mass \cite{ref.2} (measured
with an accuracy of 0.4 MeV) allows one to determine $G_2$ with 
better precision. In this paper we show why a precise  knowledge of
$\Delta_{\rm HF} (1P)$ in charmonium is so important for a fundamental
theory and explicitly extract the gluonic condensate $G_2$  from
existing experimental data.

\section{The perturbative HF interaction}
We consider here the perturbative HF interaction in one-loop
approximation which is well known. The splitting $\Delta_{\rm HF}(1P)$
is given by the expression \cite{ref.5}:
\be
 \Delta^P_{\rm HF} (1P) =\frac89 \left[ \frac14-\frac{n_f}{3}\right]
 \frac{\alpha_s^2(\mu)}{\pi} \frac{1}{m^2_q} \lan r^{-3}\ran_{1P} .
\label{eq.4}
\ee  
It is important to take the HF splitting just  in the form (\ref{eq.4})
while another (approximate) definition suggested in Ref.~\cite{ref.5},
\begin{equation}
 \Delta^P_{\rm HF} (1P) =\frac{10}{81} \left[
 \frac14-\frac{n_f}{3}\right] \frac{\alpha_s}{\pi}
 \left\{M(^3P_2)-M(^3P_0)\right\} ,
\label{eq.4a}
\end{equation}
cannot be used because it is valid only in lowest order and neglects
second order corrections $(\sim\alpha^2_s/\pi)$ and the NP
contribution. It will be  shown  later that the contribution of the
neglected terms to the difference $M(\chi_{c2})-M(\chi_{c0}) =3a +0.9
c$ (where $a$ and $c$ are the spin-orbit and the tensor splitting) is
about 26\% and has negative sign (see the numbers in
Table~\ref{table.1}).

It is also important that if relativistic corrections are taken into
account, the current (pole) quark mass in (\ref{eq.4}) must be replaced
by the average kinetic energy, $\omega_q= \lan \sqrt{\vep^2+m^2_q}
\ran_{1P}$, (usually called the constituent quark mass). This
modification can  be rigorously derived within the
Fock-Feynman-Schwinger representation of the gauge-invariant meson
Green function, when the spin-dependent interaction can be considered
as a perturbation \cite{ref.10}. Then instead of (\ref{eq.4}) one 
must use
\be
 \Delta_{\rm HF}^P(1P) = - \frac{26}{27} \frac{\alpha^2_s(\mu)}{\pi}
 \frac{1}{\omega^2_q} \lan r^{-3}\ran_{1P} \quad (n_f=4).
\label{eq.5}
\ee 
As seen from Eq.~(\ref{eq.5}) the HF splitting strongly depends on the
coupling $\alpha_s(\mu)$, where the scale $\mu$ in $\alpha_s(\mu)$
cannot be arbitrary. It is clear that for the $^1P_1$ state the scale
should be  the same as in the coupling $\alpha_{\rm FS}(\mu)$ used in
the fine structure (FS) splittings of the $\chi_c(^3P_J)$ mesons.
Fortunately, $\alpha_{\rm FS}(\mu)$ can be directly  extracted from the
experimental values of the spin-orbit and tensor splittings: as derived
in \cite{ref.11} the following relation is valid 
\be
 \alpha^2_{\rm FS} (\mu) =
 \frac{\pi\omega^2_q\{\eta_c(\exp)-|a_{\rm NP}(1P)|\}}{2f_4(1P)}.
\label{eq.6} 
\ee 
Here $\eta_c(\exp) =\frac32 c(\exp) -a(\exp)= 0.024(1)$ GeV; the tensor
splitting  $c(\exp) =0.039 (1)$  GeV  and the spin-orbit splitting
$a(\exp) =0.0346(2)$ GeV are calculated  from the $\chi_{cJ}$ masses.

In Eq.~(\ref{eq.6}) $a_{\rm NP} (1P)$ is the Thomas precession (NP) term: 
\begin{equation}
 a_{\rm NP}{(1P)} =-\frac{\sigma}{2\omega^2_q} \lan r^{-1}\ran_{1P},
\label{eq.6a}
\end{equation}
while $f_4$ is the matrix element  (m.e.) entering the second order
\underline{perturbative} part of the parameter $\eta_P$:
\be
 \eta_P=\frac32 c_P -a_P =\frac32 c_P^{(2)} - a_P^{(2)}
 =\frac{2\alpha^2_{\rm FS} (\mu)}{\pi\omega^2_q} f_4.
\label{eq.7}
\ee 
For the extraction of the strong coupling constant it is essential that
the m.e. $f_4$ does not explicitly depend on the renormalization scale $\mu$:
\be
 f_4= 1.97834\lan r^{-3}\ran_{1P} -\lan r^{-3}\ln (\bar m r)\ran_{1P}
 \quad (n_f=4).
\label{eq.8}
\ee 
The derivation of the HF interaction in Ref.~\cite{ref.5} shows that
the mass $\bar m (\bar m)$, entering $f_4$, is the current (Lagrangian)
mass $(\bar m =1.16$ GeV and $m_c$ (pole)=1.45 GeV are  taken here).

The analysis in Ref.~\cite{ref.11} has shown that the extracted value
of $\alpha_{\rm FS}(\mu_{\rm FS})$ (in the $\overline{MS}$ scheme) and
the scale $\mu_{\rm FS}$ significantly differ if the constituent mass
$\omega_c$ (about 1.6 $\div $ 1.7 GeV) instead of the pole mass $m_c$
(about 1.4 $\div 1.48$ GeV) enters the expression (\ref{eq.6}). For
example, the precise description of the FS splitting (with an
accuracy better than 1\%) for $m_c({\rm pole}) =1.45$ GeV in Eq.~(\ref{eq.6})
gives
\be
 \alpha_{\rm FS} (\mu_1) =0.358\; {\rm with}\; \mu_1=0.70\, {\rm GeV}.
\label{eq.9}
\ee 
In relativistic calculations with $\omega_c (1P) = \lan
\sqrt{\vep^2+m^2_c}\ran_{1P}\cong 1.66$ GeV the value of the coupling
constant $\alpha_{\rm FS}(\mu_R)$ extracted
from Eq.~(\ref{eq.6}) appears to be significantly larger,
\be \alpha_{\rm FS} (\mu_R) = 0.514;~~ \mu_R=0.51~{\rm GeV} .
\label{eq.10}
\ee 
From our point of view it is the second choice which should be
prefered, because the value $\alpha_{\rm FS} (\mu_1)$ in
Eq.~(\ref{eq.9}) is too small for such a small scale as $\mu_1 =0.70$
GeV. It is known that the strong coupling in the $\overline{MS}$ scheme
is already rather large at the scale $M_\tau=1.777$ GeV
$(\alpha_s(M_\tau)\ga 0.33$ \cite{ref.3}), while at the scale $\mu_{\rm
FS} \cong 1.0$ GeV the coupling $\alpha_{\rm FS} (1.0$ GeV)$\cong 0.40$
has been obtained in \cite{ref.12} in the analysis of FS splittings of
the 2$P$ state in bottomonium. Therefore the values given in Eq.~
(\ref{eq.10}) will be taken here. They result in the following FS
splittings of the $\chi_{cJ}$ mesons (with an accuracy better than 1\%):
\be
\begin{array}{ll} c(1P)=39.12{\rm MeV},& c_{\exp}(1P) =39.12\pm 0.62 {\rm MeV},
\\ 
 a(1P)=34.58{\rm MeV},& a_{\exp}(1P) =34.56\pm 0.19 {\rm MeV}.
\end{array} 
\label{eq.11}
\ee
The first- and second-order perturbative and NP terms in the FS
splittings:  $a(1P) =a^{(1)}_p + a_p^{(2)}+ a_{\rm NP}$, $c(1P)
=c_p^{(1)}+c_p^{(2)}$ $ (c_{\rm NP}$ is very small \cite{ref.11}), are
given in Table~\ref{table.1}.

\begin{table}
\begin{center}
\caption{\label{table.1}The spin-orbit and tensor splittings (in MeV)
of the $\chi_{cJ}$ mesons (the static potential is  taken from
\cite{ref.13} with the parameters  $m_c$(pole)$=1.45$ GeV,
$\omega_c=1.66$ GeV, $\sigma=0.18$ GeV$^2$,
$\Lambda^{(4)}_{\overline{MS}}$(2-loop)$=267$ MeV, $\alpha_{\rm
FS}(\mu_{\rm FS}=0.51$ GeV)=0.514.}
\begin{tabular}{|l|l|l|l|}
\hline
  & ~1st order term& ~2d order term & ~NP term \\
\hline
 ~$a$(tot)$=34.58$& ~$a_p^{(1)}=51.92~$ & ~$a_p^{(2)}=-4.24$~& ~$a_{\rm NP}
  =-13.10$~\\
 ~$a(\exp)=34.56(19)~$&&&\\ 
\hline
 ~$c$(tot)$=39.12$~& ~$c_p^{(1)}=34.61$~& ~$c_p^{(2)}=4.51$~&
 ~$c_{\rm NP}< 1$ MeV$^{a)}$~\\
 ~$c(\exp)=39.12(62)$~&&&\\
\hline
\end{tabular}
\end{center}

$^{a)}$ The reasons why $c_{\rm NP} (1P)$ is small, are discussed in
\cite{ref.11,ref.12}.
\end{table}

Having obtained a  precise description of the FS of the  $\chi_{cJ}$
mesons, we can expect that the HF splitting (for the same set of
physical parameters) is also determined with good accuracy. From the
expression (\ref{eq.5}) (with $\omega_c=1.66$ GeV, $\alpha_{\rm
HF}(\mu_R)=\alpha_{\rm FS} (\mu_R=0.51 $ GeV)$ =0.514,~\lan
r^{-3}\ran=0.139)$ it follows that in charmonium the perturbative
contribution has the value
\be
\Delta_{\rm HF}^{\rm P} (1P) =- 4.1 ~{\rm MeV},
\label{eq.12}
\ee
which is five times larger than the one obtained in the experiments
\cite{ref.1,ref.2} (in the CLEO experiment \cite{ref.1} $\Delta_{\rm
HF}$ is positive, 1.0 $\pm$1.0 MeV).

The matrix elements Eq.~(\ref{eq.8}) are calculated here utilizing the
solutions of the spinless Salpeter equation with a static
potential--linear plus  gluon-exchange term, where in two-loop vector
coupling the asymptotic freedom behavior at small distances and the
freezing of the coupling at large distances are taken into account.
The most important matrix elements are $\lan r^{-1}\ran =0.405$ GeV, $\lan
r^{-3}\ran_{1P} =0.139$ GeV$^3$, and $\lan r^{-3} \ln (\bar m r)\ran
=0.095$ GeV$^3$.  Note that the m.e.$\lan r^{-3}\ran$ in the
relativistic case is about 30\% larger than in nonrelativistic
calculations.

\section{The nonperturbative HF interaction}

Spin-dependent NP potentials have been introduced in
\cite{ref.14,ref.15}. With the use of the vacuum correlation function
(v.c.f.) $D(x)$ the  HF interaction is written as
\be
 V_{\rm HF}^{\rm NP} =\frac{1}{3\omega^2_c} V_4^{\rm NP} (r)
\label{eq.13}
\ee
with
\be
 V_4^{\rm NP} (r) =6 \int^\infty_0 d\nu D(\sqrt{r^2+\nu^2}) .
\label{eq.14}
\ee
The contribution of the other correlator, $D_1(x)$, has been neglected
in Eq.~(\ref{eq.14}), because in the unquenched case $D_1(x)$ is small,
even compatible with zero \cite{ref.16}.  The v.c.f. $D(x)$ is
shown to behave as an exponential: $D(x) =d\exp (-x/ T_g)$ at $x\ga
0.2$ fm  \cite{ref.16,ref.17} ($T_g$ is the gluonic correlation
length), while at smaller $x<r_0$ ($r_0\la 0.2$ fm) $D(x)$  should have
a plateau to satisfy the necessary conditions
$\left.\frac{d^2D(x)}{dx^2}\right|_{x= 0} <0$ and
$\left.\frac{dD}{dx}\right|_{x= 0}  =0$, established in \cite{ref.18}.

The value of $D(x)$ at the origin, $D(0)$ is related to the gluonic
condensate $G_2=\frac{\alpha_s}{\pi} \lan  F_{\mu\nu} (0) F_{\mu\nu}
(0)\ran$:
\be
 D(0)=\frac{\pi^2}{18} G_2,
\label{eq.15}
\ee  
and we assume that $D(r_0) \cong D(0)$, so the factor $d$ in front of
the exponent is $d=D(0)\exp (r_0/ T_g).$
At the origin the HF interaction Eq.~(\ref{eq.13}) has the value
$V_{\rm HF}^{\rm NP} (r = 0) = \pi^2 G_2\,(r_0 + T_g)/(9 \omega^2_c)$.

The string tension in the confining potential is defined through
the same v.c.f. $D(x)$: 
\be
 \sigma= 2 \int^\infty_0 d\lambda \int^\infty_0 d\nu D(\sqrt{\lambda^2+\nu^2})
\label{eq.16}
\ee 
and for the form adopted for $D(x)$ we find
\be
 \sigma=\frac{\pi^3}{18} G_2 T^2_g \left[1 + \frac{r_0}{T_g}
 + \frac{1}{2} \left(\frac{r_0}{T_g}\right)^2 \right].
\label{eq.17}
\ee
From  this relation and taking the ``physical'' value $\sigma\approx
0.178(8)$ GeV$^2$ one can determine the gluonic condensate for
different values of $T_g$. However, the gluonic correlation length is
not known with good accuracy and at present different values (between
0.16 fm and  0.26 fm) have been obtained in lattice QCD
\cite{ref.17,ref.16} and in the vacuum correlator method
\cite{ref.19}.  For comparison  we use  here two values, $T_g=0.16$ fm
from Ref.~\cite{ref.19} and 0.2 fm. We do not consider here the value
$T_g \approx 0.3$ fm, because our analysis shows that for such a value
of $T_g$ the NP contribution to the HF splitting is too large
($\Delta^{\rm NP}_{\rm HF} (1P) \approx 10$ MeV) so that the total HF
splitting is positive, about 6 MeV, i.e. several times larger than the
experimental value given in Eq.~(\ref{eq.2}).

Then  from the relation (\ref{eq.17}) with $\sigma =0.178(8)$
GeV$^2$ it follows that 
\beq
 G_2=0.065(3)\; {\rm GeV}^4 & & (r_0 = T_g=0.16\;{\rm fm}), \nonumber 
\\ 
 G_2=0.041(2)\; {\rm GeV}^4 & & (r_0 = T_g=0.20\;{\rm fm}).
\label{eq.18}
\eeq

The HF splitting, calculated  for the  interaction (\ref{eq.14}),
reduces to the expression \cite{ref.7},
\be
 \Delta_{\rm HF}^{\rm NP} (1P) =
 \frac{\pi^2}{9} \frac{G_2}{\omega_c^2} (r_0 + T_g) J,
 \quad J={\left \lan rK_1 \left(\frac{r}{T_g}\right)\right\ran}_{1P}.
\label{eq.19}
\ee 
The accuracy of this approximation is better than 5\%. The matrix
element $J$, $J(T_g=0.16\,{\rm fm})=0.092$ GeV$^{-1}$ and
$J(T_g=0.20\,{\rm fm})=0.17$ GeV$^{-1}$,  strongly depends on $T_g$.
Then taking $G_2$ from (\ref{eq.18}) one obtains
\beq
 \Delta^{\rm NP}_{\rm HF} (1P) = 3.8(4)\; {\rm MeV}& &
 (T_g=0.16\;{\rm fm}),
\nonumber\\
 \Delta^{\rm NP}_{\rm HF} (1P) = 5.5(3)\; {\rm MeV}& & (T_g=0.20\;{\rm fm}).
\label{eq.20}
\eeq

Thus the magnitude of the NP contribution (\ref{eq.20}) appears to be
larger  (smaller) than the perturbative term (\ref{eq.12}) for larger
(smaller) gluonic correlation length. Therefore the total HF splitting
Eq.~(\ref{eq.3}) has different signs for $T_g=0.16$ fm and $T_g=0.2$ fm,
\beq
 \Delta_{\rm HF}(1P) = -0.3(4)\; {\rm MeV} & & (T_g=0.16\;{\rm fm})
\label{eq.21} \\
\Delta_{\rm HF} (1P) = +1.4(2)\; {\rm MeV} & & (T_g=0.20\;{\rm fm}).
\label{eq.22} 
\eeq

It is amusing to notice that the HF splitting for $T_g=0.16$ fm exactly
coincides with the experimental value obtained in the E835 experiment
\cite{ref.2}, while the positive splitting (\ref{eq.22}) is close to
the value obtained in the CLEO experiment \cite{ref.1}.  To distinguish
between these two possibilities $\Delta_{\rm HF} (1P)$ needs to be
measured in charmonium with a better accuracy.

\section{Conclusions}

Our analysis has shown that the ``physical'' value of the string
tension cannot unambigously fix the gluonic condensate and only
the product $G_2T^2_g$ can be determined.

The HF splitting between the c.o.g of the $\chi_{cJ}$ mesons and
$h_c(^1P_1)$ provides an additional opportunity to extract the
gluonic condensate with good accuracy, due to the cancellation
between the negative perturbative and positive NP contributions which
both have small magnitude.

To calculate the HF splitting for the 1$P$ states in charmonium we use
exactly the same coupling $\alpha_s (\mu)$ and matrix elements as in
the fine structure analysis of the $\chi_{cJ}$ mesons where a high
accuracy  ($\la 1\%$) is reached. Therefore we estimate the accuracy of
our calculations of $\Delta_{\rm HF}(1P)$ to be equal to 0.3 MeV (0.4
MeV) for the gluonic correlation length $T_g=0.16$ fm $(T_g=0.2$ fm).

An additional restriction is also put on the gluonic condesate--it
should correspond to the ``physical'' string tension, $\sigma\cong 0.18$
GeV$^2$, used in the static potential.

Then a negative central value of $\Delta_{\rm HF}(1P) = -0.3$ MeV, as
in the E835 experiment  \cite{ref.2}, is obtained for $G_2 = 0.065$
GeV$^4 (T_g=0.16$ fm). For the larger correlation length, $T_g=0.2$ fm,
and the smaller value $G_2 =0.041$ GeV$^4$ the HF splitting appears to
be  positive, $\Delta_{\rm HF} (1P) = +1.4$ MeV.
So it is of prime importance for the deduction of the values of the
gluonic condensate and correlation length that the discrepancy between
the two experimental results be removed.

\acknowledgments
We thank Yu.A. Simonov for fruitful discussions.


\begin{thebibliography}{99}

\bibitem{ref.1} A. Tomaradze, hep-ex/0410090.

\bibitem{ref.2}  C. Patrignani, hep-ex/0410085;
T.A. Armstrong et al., Phys. Rev. Lett. {\bf 69}, 2337 (1992).

\bibitem{ref.3} Particle Data Group, K. Hagiwara et al., 
Phys. Rev. D {\bf 66}, 010001 (2002).

\bibitem{ref.4} A.M. Badalian and V.P. Yurov, 
Phys. Rev. D {\bf 42}, 3138 (1990).

\bibitem{ref.5} K. Igi and S. Ono,
Phys. Rev. D {\bf 33}, 3349 (1986); Errata: Phys. Rev. D {\bf 37}, 1338 (1987);
J.~Pantaleone, S.-H.H. Tye, and Y.J.~Ng, Phys. Rev. D {\bf 33}, 777 (1986).

\bibitem{ref.6} S. Titard and F.J.~Yndurain,  
Phys. Lett. B {\bf 351}, 541 (1995).

\bibitem{ref.7} A.M. Badalian and V.L.~Morgunov, 
Phys. Atom. Nucl. {\bf 62}, 1019 (1999); Yad. Fiz. {\bf 62}, 1086 (1999).

\bibitem{ref.8} A.I. Vainshtein, V.I. Zakharov, and M. Shifman, 
JETP Letters, {\bf 27}, 55 (1978); Nucl. Phys. B {\bf 165}, 45 (1980);
V.I. Zakharov, Int. J. Mod. Phys. A {\bf 14}, 4865 (1999);
B.V. Geshkenbein, Phys. Rev. D {\bf 70}, 074027 (2004);
S. Narison hep-ph/0411145.

\bibitem{ref.9} H.G. Dosch, M. Eidenmuller, and M. Jamin, 
Phys. Lett. B {\bf 452}, 379 (1999).

\bibitem{ref.10} Yu.A.Simonov, 
in Proceedings of the XVII International Sch. Physics, Lisbon, 1999, 
edited by L.S. Ferreira, P. Nogueira, and J.L. Silva-Marcos
(World Scientific, River Edge, 2000), p~60

\bibitem{ref.11} A.M. Badalian and V.L. Morgunov, 
Phys. Rev. D {\bf 60}, 116008 (1999).

\bibitem{ref.12} A.M. Badalian and B.L.G. Bakker,
Phys. Rev. D {\bf 62}, 094031 (2000).

\bibitem{ref.13} A.M. Badalian, B.L.G. Bakker, and A.I. Veselov, 
Phys. Rev. D {\bf 70}, 016007 (2004); 
Phys. Atom. Nucl. {\bf 67}, 1367 (2004); Yad. Fiz. {\bf 67}, 1392 (2004).

\bibitem{ref.14} E. Eichten and F.L. Feinberg, 
Phys. Rev. D {\bf 23}, 2724 (1981).

\bibitem{ref.15} A.M. Badalian and Yu.A. Simonov, 
Phys. Atom. Nucl. {\bf 59}, 2164 (1996); Yad. Fiz. {\bf 59}, 2247 (1996).

\bibitem{ref.16} M. D'Elia, A. Di Giacomo, and E. Meggiolaro, 
Phys. Lett. B {\bf 408}, 315 (1997);
A. Di Giacomo and H. Panagopoulos, Phys. Lett. B {\bf 285}, 133 (1992).

\bibitem{ref.17} G.S. Bali, N. Brambilla, and A. Vairo, 
Phys. Lett. B {\bf 421}, 265 (1998).

\bibitem{ref.18} Yu.A. Simonov, 
Phys. Atom. Nucl. {\bf 50}, 310 (1989); Yad. Fiz. {\bf 50}, 213 (1989);
V.I. Shevchenko and Yu.A. Simonov, 
Phys. Atom. Nucl. {\bf 60}, 1201 (1997); Yad. Fiz. {\bf 60}, 1329 (1997).

\bibitem{ref.19} Yu.A. Simonov, Nucl. Phys. B {\bf 592}, 350 (2001).

\end{thebibliography}
\end{document}